\documentclass[fleqn,usenatbib,useAMS]{mnras}

\usepackage{graphicx}	
\usepackage{amsmath}	
\usepackage{amssymb}	
\usepackage{amsfonts}
\usepackage{amsbsy}
\usepackage{color}
\usepackage{rotating}
\usepackage{multicol}        
\usepackage{dcolumn}
\usepackage{bm}		
\usepackage{lscape}	
\usepackage[T1]{fontenc}
\usepackage[english]{babel}
\usepackage{ae,aecompl}

\title {{Clustering of Galaxies with $f(R)$ gravity}}
\author[S. Capozziello et al.]{
Salvatore Capozziello,$^{1,2,3,4}$\thanks{E-mail: capozzie@na.infn.it}
Mir Faizal,$^{5,6}$\thanks{E-mail: mirfaizalmir@googlemail.com}
Mir  Hameeda,$^{7,8}$\thanks{E-mail: hme123eda@gmail.com}
Behnam Pourhassan,$^{9}$\thanks{E-mail: b.pourhassan@du.ac.ir}
\newauthor Vincenzo Salzano$^{10}$\thanks{E-mail: vincenzo.salzano@usz.edu.pl}
Sudhaker Upadhyay,$^{11}$\thanks{E-mail: sudhakerupadhyay@gmail.com}
\\
$^{1}$Dipartimento di Fisica ``E. Pancini",  Universit\'a di Napoli  Federico II, I-80126 - Napoli, Italy \\
$^{2}$INFN Sez. di Napoli, Compl. Univ. di Monte S. Angelo, Edificio G, I-80126 - Napoli, Italy \\
$^{3}$Gran Sasso Science Institute (INFN), Viale F. Crispi, 7, I-67100, L'Aquila, Italy \\
$^{4}$ Tomsk State Pedagogical University, ul. Kievskaya, 60, 634061 Tomsk, Russia\\
$^{5}$Irving K. Barber School of Arts and Sciences, University of British Columbia - Okanagan, 3333 University Way, Kelowna, \\ British Columbia V1V 1V7, Canada\\
$^{6}$ Department of Physics and Astronomy, University of Lethbridge, Lethbridge, Alberta, T1K 3M4, Canada \\
$^{7}$Department of Physics, S.P. Collage,  Srinagar, Kashmir, 190001, India\\
$^{8}$Visiting Associate, IUCCA,  Pune,  41100, India\\
$^{9}$ School of Physics, Damghan University, Damghan, 3671641167, Iran \\
$^{10}$ Institute of Physics, Faculty of Mathematics and Physics, University of Szczecin, Wielkopolska 15, 70-451 Szczecin, Poland\\
$^{11}$Centre for Theoretical Studies, Indian Institute of Technology Kharagpur,  Kharagpur-721302,  India}

\date{Accepted XXX. Received YYY; in original form ZZZ}

\pubyear{2017}

\begin{document}

\label{firstpage}
\pagerange{\pageref{firstpage}--\pageref{lastpage}}
\maketitle

\begin{abstract}
Based on thermodynamics, we discuss the galactic clustering of  expanding Universe by
assuming the  gravitational interaction through the modified  Newton's potential given by $f(R)$ gravity.  We compute the corrected $N$-particle partition function analytically. The corrected partition function leads to more exact equations of states of the system.  By assuming that system follows quasi-equilibrium,  we derive the exact distribution function which exhibits
the $f(R)$ correction. Moreover, we evaluate the critical temperature and discuss the stability of the system. We observe the effects of correction of $f(R)$ gravity on the power law behavior of particle-particle correlation function also. In order to check
feasibility of an $f(R)$ gravity approach to
the clustering of galaxies,  we compare our results with an observational galaxy cluster catalog.
\end{abstract}


\section{Overview and Motivations}
The distribution of galaxies is influenced by the gravitational force \citep{pad}. In fact, gravitational force
 plays an important role   not only in clustering of galaxies but also  in their large scale structure formation. The characterization of galactic clusters on very large
scales is very crucial to understand  the evolution and
distribution of the galaxies throughout the Universe.
 One of the standard ways to study
the formation of the Universe is study of the
correlation functions by means of observation \citep{pee80} and  $N$-body computer
simulations \citep{itoh}.  It is known that the gravitating systems, which interact  in pairs,
the  correlation functions determine the thermodynamical properties which includes gravitation \citep{hill}.

An empirically motivated modification of Newtonian dynamics, so-called  MOND, has been proposed at low accelerations  \citep{mil},
which had explained successfully a number of observational properties of galaxies \citep{san}.
In MOND scenario,  the modified gravitational
potential  in the weak field limit of $f(R)$ gravity has estimated the total mass of a sample of 12 clusters of galaxies which  also
fits to the mass of visible matter estimated by X-ray observations \citep{2aa}.
The effects of modified gravitational potential on clustering of galaxies have been studied recently.
For instant, the effects of dark energy on galactic clustering is studied in Refs. \citep{MNRAS,new}. Recently, the galactic clustering of an expanding Universe by considering the
logarithmic and volume (quantum) corrections to Newton's law along with the repulsive effect of a harmonic
force induced by the cosmological constant  is also discussed \citep{sud}.
The clustering of galaxies under brane world modified gravity is also analysed recently \citep{ham}.
Our motivation here is to discuss the effects of $f(R)$ gravity on the galactic clustering
and compare our results   with the observations. For details on these theories see, for example  \cite{Odintsov,Review,annalen,Oikonomou}.

In this paper, we first calculate the modification in the Newtonian potential
due to the $f(R)$ gravity.  The modified Newtonian potential leads to an explicit form of
$N$-particle (galaxies) configuration integrals which helps us to estimate the $N$-particle
partition function.  Once the partition function is known, it is matter of calculation to
derive various important thermodynamical entities. For instance, we derive
Helmholtz free energy, entropy, internal energy, pressure and chemical potential.
Remarkably, the corrected form of clustering parameter emerges naturally from these thermodynamical quantities. With the help of this clustering parameter and generating functional, we derive the exact expression for  distribution function  for both point and non-point galaxies and study the effects of correction term  on the distribution function.
Furthermore, we demonstrate the corrected specific heat at constant volume and by maximizing
it we compute critical temperature for the gravitating system under $f(R)$ gravity. Remarkably,
we obtain the same form of corrected specific heat in terms of critical temperature
to that of without  correction. Moreover, we analyse the evolution of galaxy-galaxy correlation function where the effects of $f(R)$ modifications are evident. Finally, we have also tried to test the proposed $f(R)$ gravity model by the obtained distribution function with observational data. While we obtain a very good description of the observed clustering of galaxies with our formulas, we have also to stress that we cannot constrain in a statistically significant way most of the $f(R)$ parameters, due to the large degeneracies among them.

The plan of the paper is as following.
In section 2, we discuss $f(R)$ gravity and the $N$ particle  partition function of
the system under the gravitational potential of an analytic $f(R)$ theory.
In section 3, we study   various thermodynamical equation of states   relevant for strongly
interacting system of galaxies interacting through the modified Newtonian potential
due to $f(R)$ gravity. In section 4, we estimate more exact distribution function for
such gravitating system. The critical temperature is derived in section 5. The
  power law behavior for the galaxy-galaxy correlation function under $f(R)$ modified
Newtonian potential is discussed in section 6. The theory is matched with observations
in section 7. Finally, we draw conclusions and make final remarks in the last section.

\section{Gravitational Partition Function}
Let us first write the action for $f(R)$ gravity as follows \citep{2aa},
\begin{eqnarray}
S =\frac{1}{16\pi G}\int d^4x \sqrt{-g}\left[f(R) +{\cal L}_{matter}\right],
\end{eqnarray}
where  $f(R)$ is an analytic function of the Ricci scalar $R$, $g$ refers the determinant of the metric $g_{\mu\nu}$ and ${\cal L}_{matter}$ is the standard Lagrangian for the  perfect fluid matter. The   equations of motion corresponding to the metric is  given by
\begin{eqnarray}
 f'R_{\mu\nu}-\frac{1}{2}g_{\mu\nu}f'_{;\mu\nu}+g_{\mu\nu}\square f'= {8\pi G} T_{\mu\nu},
\end{eqnarray}
where the prime indicates the derivative with respect to $R$. Here  $T_{\mu\nu}$ is the energy momentum tensor of matter given as
$T_{\mu\nu}=\frac{-2}{\sqrt{-g}}\frac{\delta(\sqrt{-g}{\cal L}_{matter}) }{\delta g^{\mu\nu}}$. In order to study the most general result, we only consider analytic Taylor expandable functions
\begin{eqnarray}
f(R)\simeq f_0+f_1 R+f_2 R^2+.... .
\end{eqnarray}
Here $f_1$ and $f_2$ are the expansion coefficients and dots include higher order terms \citep{Huang} and also logarithmic corrected term \citep{Sadeghi}.

In order to discuss the physical prescription of the asymptotic flatness at infinity, the metric becomes \citep{2aa},
\begin{eqnarray}
ds^2&=&\left[1-\frac{2GM}{f_1r} -\frac{ \delta_1(t)e^{-r\sqrt{-\xi}}}{3\xi r}
-\frac{\Lambda r^2}{3}\right]dt^2\nonumber\\
&-&\left[1+\frac{2GM}{f_1r} -\frac{ \delta_1(t)(r\sqrt{-\xi}+1)e^{-r\sqrt{-\xi}}}{3\xi r}
 \right]dr^2\nonumber\\
&-&r^2d\Omega,\label{cc}
\end{eqnarray}
where  the integration  constant $\delta_1(t)$ is arbitrary time-function and  {has the dimensions of $length/time^2$}, and $d\Omega$  is the angular element. Here $\xi=\frac{f_1}{6f_2}$ has the dimension of $1/(length)^2$,  {with $f_{1}$ being dimensionless, and $f_{2}$ having the dimension of $(length)^2$}.

As we have an explicit Newtonian-like term into the definition, the solution can be given   in terms of gravitational potential $U(r)$.  In particular, it is $g_{tt} = 1+2U(r) $ and then from (\ref{cc}) the gravitational potential of an analytic $f(R)$-theory given by,
\begin{eqnarray}
U(r)= -\frac{GM}{f_1r} -\frac{\delta_1(t)e^{-r\sqrt{-\xi}}}{6\xi r},
\end{eqnarray}
where $\Lambda=0$ assumed. However it is possible to consider cosmological constant $\Lambda$ as dark energy to find the effect of dark energy on the cluster of galaxies \citep{MNRAS}. In that case the effect of dynamical dark energy on the cluster of galaxies \citep{new} can be modeled using varying $\Lambda$ or $G$ \citep{var}.
Gravitational potential (\ref{cc}) corresponds  to the following potential energy,
\begin{eqnarray}
\Phi(r)=-\frac{GM^2}{f_1r} -\frac{M\delta_1(t)e^{-r\sqrt{-\xi}}}{6\xi r}.\label{ph}
\end{eqnarray}

The thermodynamics applies as a description of various dynamical system on a macroscopic level
with  both equilibrium and non-equilibrium states. The argument for validity for
non-equilibrium systems is that the globally averaged thermodynamical quantities like
pressure, temperature, density and internal energy change more slowly than completely
specified local configurations of particles. In gravitating $N$-body systems
no rigorous equilibrium is possible because the gravitation is a long range force and does not saturate.
Thus, the  system of galaxies  can be approximated as particles with  pairwise interaction.

In order to study the thermodynamics of $N$  particles or galaxies with equal mass $M$ interacting gravitationally with a potential energy
$\Phi(r)$ given by (\ref{ph}),   momenta $p_{i}$ and average temperature $T$,
the   partition function is given by,
\begin{eqnarray}
&&Z(T,V)=\frac{1}{\lambda^{3N}N!}\int d^{3N}pd^{3N}r \times\nonumber\\
&& \exp\biggl(-\biggl[\sum_{i=1}^{N}\frac{p_{i}^2}{2M}+\Phi(r)\biggr]T^{-1}\biggr),
\end{eqnarray}
where $N!$ corresponds to the distinguishability of classical particles, and $\lambda$
takes care of
the normalization factor  resulting from integration over momentum space.
Upon  integration over momentum space, this reduces to
\begin{eqnarray}
Z_N(T,V)=\frac{1}{N!}\left(\frac{2\pi MT}{\lambda^2}\right)^{3N/2}Q_N(T,V), \label{zn}
\end{eqnarray}
where the configurational integral, $Q_{N}(T,V)$, is  given by
\begin{equation}
Q_{N}(T,V)=\int....\int \prod_{1\le i<j\le N} \exp[-\frac{\phi(r_{ij} )}{T}]d^{3N}r. \label{q1}
\end{equation}
Here, the sum of the potential energies of all pairs  is a function of the relative position vector $r_{ij}=|r_{i}-r_{j}|$ and corresponds  $\sum_{1\le i<j\le N}\phi(r_{ij})= \Phi(r)$.

The configurational integral in terms of two particles function can be expressed as  \citep{ahm02}
\begin{eqnarray}
Q_{N}(T,V)&=&\int....\int \biggl[(1+f_{12})(1+f_{13})(1+f_{23})\nonumber\\
&& \dots (1+f_{N-1,N})
\biggr]d^{3}r_{1}d^{3}r_{2}\dots d^{3}r_{N},\label{q2}
\end{eqnarray}
where two particles function has following form:
\begin{equation}
f_{ij}=e^{-\Phi(r_{ij})/T}-1. \label{fun}
\end{equation}
Here we note that the appearance of two particles function
confirms the presence of interactions in the system and vanishes   in
absence of interactions.

At the cosmological scale we consider galaxies as a point particles,
and in this approximation, the Hamiltonian and, therefore, the partition function
of the systems diverge  at $r_{ij}=0$. In order to overcome this divergence,
 we consider the  extended nature of particles (for example theory of infinitely extended particles of \citep{Hessabi} which reconsidered recently \citep{MNRAS}) by introducing a softening parameter which takes care of the finite size of each galaxy.

By introducing the softening parameter $\epsilon$, the $f(R)$ modified  gravitational (interaction) potential energy between particles  which given by the equation (\ref{ph}) becomes
\begin{equation} \label{12}
\Phi(r)=-\frac{GM^2}{f_1(r^{2}+\epsilon^{2})^{1/2} } -\frac{M\delta_1(t)e^{-r\sqrt{-\xi}}}{6\xi (r^{2}+\epsilon^{2})^{1/2} }.
\end{equation}

The   two particle function  corresponding to the potential energy (\ref{12})
has the following form:
\begin{equation}
f_{ij}(r)=\exp\biggl[\frac{GM^2}{f_1(r^{2}+\epsilon^{2})^{1/2}  T}+\frac{M\delta_1(t)e^{-r\sqrt{-\xi}}}{6\xi (r^{2}+\epsilon^{2})^{1/2} T}
 \biggr]-1.
\end{equation}
 As the system is  moderately dilute, we can ignore higher order terms of the  two particle function $f_{ij}$ as
\begin{eqnarray}
f_{ij}(r)=\biggl[\frac{GM^2}{f_1(r^{2}+\epsilon^{2})^{1/2}  T}+\frac{M\delta_1(t)e^{-r\sqrt{-\xi}}}{6\xi (r^{2}+\epsilon^{2})^{1/2} T}
 \biggr].
\end{eqnarray}
The configuration integrals over a spherical volume of radius $R_{1}$ for $N=1$
yields,
\begin{equation}
Q_{1}(T,V)=V.
\end{equation}
Moreover, the configuration integral $Q_{2}(T,V)$ for $N=2$  has the following form:
\begin{eqnarray}
Q_{2}(T,V)&=&4\pi V \int_{0}^{R_{1}}\left[1+\frac{GM^2}{f_1(r^{2}+\epsilon^{2})^{1/2} T}
\right.\nonumber\\
&+&\left. \frac{M\delta_1(t)e^{-r\sqrt{-\xi}}}{6\xi (r^{2}+\epsilon^{2})^{1/2}T}        \right]r^2dr.
\end{eqnarray}
This further simplifies to
\begin{eqnarray}
Q_{2}(T,V)&=&V^2\left[1
+ \frac{3}{2}\frac{GM^2}{R_1T}\left(\frac{1}{f_1}\sqrt{1+\frac{\epsilon^2}{R_1^2}}\right.\right.\\
&+&\left.\left.  \frac{\epsilon^2}{f_1R_1^2} \log \frac{\epsilon/R_1}{\left[ 1+\sqrt{1+\frac{\epsilon^2}{R_1^2}}\right]} \right.\right.\nonumber\\
&+&\left.\left. \frac{\delta_1(t)}{147 GM\xi} \left[\sqrt{1+\frac{\epsilon^2}{R_1^2}}\left( -16 R_1\sqrt{-\xi} \right.\right.\right.\right.\nonumber\\
& +&\left.\left.\left.\left.  32 \frac{\epsilon^2}{R_1}\sqrt{-\xi}\left(1- \frac{\epsilon/R_1}{\left[ 1+\sqrt{1+\frac{\epsilon^2}{R_1^2}}\right]} \right) \right.\right.\right.\right.\nonumber\\
& -&\left.\left.\left.\left. 6R_1^2\xi +3(8+3\epsilon^2 \xi)
\right) \right.\right.\right.\nonumber\\
& +&\left.\left.\left. 3\frac{\epsilon^2}{R_1^2}(8+3\epsilon^2 \xi) \log \frac{\epsilon/R_1}{\left[ 1+\sqrt{1+\frac{\epsilon^2}{R_1^2}}\right]}     \right]\right) \right]. \nonumber
\end{eqnarray}
If we define the dimensional quantities as
\begin{eqnarray}\label{alp}
 \alpha_1  &=& \sqrt{1+\frac{\epsilon^2}{R_1^2}} +\frac{\epsilon^2}{R_1^2} \log \frac{{\epsilon/R_1}}{\left[ 1+\sqrt{1+\frac{\epsilon^2}{R_1^2}}\right]},\nonumber\\
   \alpha_2 &=& \frac{\delta_1(t)}{147 GM\xi} \left[\sqrt{1+\frac{\epsilon^2}{R_1^2}}\left( -16 R_1\sqrt{-\xi} \right.\right. \nonumber\\
& +&\left.\left.   32 \frac{\epsilon^2}{R_1}\sqrt{-\xi}\left(1- \frac{\epsilon/R_1}{\left[ 1+\sqrt{1+\frac{\epsilon^2}{R_1^2}}\right]} \right) \right.\right. \nonumber\\
& -&\left.\left.  6R_1^2\xi +3(8+3\epsilon^2 \xi)
\right) \right. \nonumber\\
& +&\left.  3\frac{\epsilon^2}{R_1^2}(8+3\epsilon^2 \xi) \log \frac{\epsilon/R_1}{\left[ 1+\sqrt{1+\frac{\epsilon^2}{R_1^2}}\right]}     \right],
\end{eqnarray}
then,   $Q_{2}(T,V)$ reduces to following compact form:
\begin{equation}
Q_{2}(T,V)=V^2\left[1+ \frac{3}{2}\left(\frac{\alpha_1}{f_1}+ {\alpha_2} \right)\left(\frac{ GM^{2}}{ R_{1}T}\right)^3\right].\label{a}
\end{equation}
The following scale transformations of quantity:  $\rho\rightarrow \lambda^{-3}\rho, T\rightarrow\lambda^{-1}T$ and $R_1\rightarrow \lambda R_1$  leads to
 $\frac{GM^2}{R_1T}\rightarrow (\frac{GM^2}{R_1T})^3 =\frac{3}{2}\left(\frac{ GM^{2}}{  T}\right)^3\rho \equiv x$.
With this,  equation (\ref{a}) reduces to
\begin{equation}
Q_{2}(T,V)=V^2\big(1+\alpha x\big),
\end{equation}
where
\begin{equation}
\alpha\left(\frac{\epsilon}{R_1}\right) =\frac{\alpha_1}{f_1}+ {\alpha_2}.
\end{equation}
For the point masses  (i.e., $\epsilon =0$), $\alpha$ reduces to
\begin{equation}
\alpha\left( \epsilon =0\right) =  1   +\alpha_2|_{\epsilon=0},
\end{equation}
where
\begin{eqnarray}
\alpha_2|_{\epsilon=0}= \frac{\delta_1(t)}{147 GM\xi} \left[ \left( -16 R_1\sqrt{-\xi} - 6R_1^2\xi +24
\right)    \right].
\end{eqnarray}
In same fashion,  the configurational integrals for $N=3,4$ are obtained iteratively as
\begin{equation}
Q_{3}(T,V)=V^3\big(1+\alpha x\big)^{2},
\end{equation}
and
\begin{equation}
Q_{4}(T,V)=V^4\big(1+\alpha x\big)^{3}.
\end{equation}
This enables us to write the most general form of  configurational integral as
\begin{equation}
Q_{N}(T,V)=V^N\big(1+\alpha x\big)^{N-1}.\label{qn}
\end{equation}
Exploiting definition (\ref{zn}), we can write explicit form of the gravitational partition function for gravitating system under $f(R)$ gravity as
\begin{equation}
Z_N(T,V)=\frac{1}{N!}\left(\frac{2\pi MT}{\lambda^2}\right)^{3N/2}V^{N}\big(1+\alpha x\big)^{N-1}.
\end{equation}
Here,  the correction due to $f(R)$ term is embedded in parameter $\alpha$.

The possibility to eventually estimate the constraints for $f(R)$ theories which are in literature on the parameter $\alpha$ is made difficult by many factors. First of all, in a background analysis we generally don't need to define parameters like $\delta_1$, which is functional to the gravitational potential, but not to other cosmological quantities. The departure from general relativity is generally ascribed to the parameter $f_{R} \equiv df/dR$ which is almost equivalent to our $f_1$, stated that most of the constraints on $f_R$ are obtained assuming a model with $R+f(R)$. Thus, our $f_{1}\sim1$ would correspond to $f_R\sim 0$. One of the latest analysis of one of the most used models in literature, the Hu-Sawicki model \citep{HuSaw}, gives an upper limit $|f_R| \lesssim 10^{-3}$, using dynamical probes (growth of matter perturbations and CMB power spectra data, among others) \citep{Hu}. This means that $f_1\lesssim 1$ (because $f_R$ from that analysis is negative) and, thus, makes a negligible contribution to $\alpha$ through the term $\alpha_1/f_1$.

Unfortunately, we cannot investigate the influence on $\alpha$ from $\delta_1$ because the potential written as in Eq.~(\ref{12}) has not been studied extensively in literature. A similar, but not equivalent version of it, can be found in \citep{2aa}, but also in this case the parameter $\delta_1$ was not explicitly left free in the analysis.

\section{Equations of State}
\begin{figure}
\centering
\includegraphics[width=0.4\textwidth]{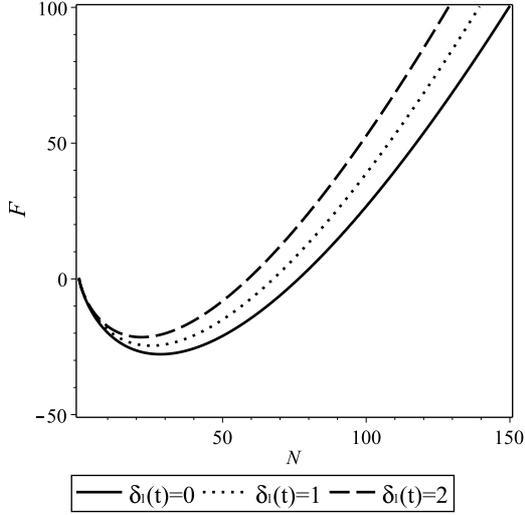}\\
~~~\\
~~~\\
~~~\\
~~~\\
~~~\\
\caption{Typical behavior of the Helmholtz free energy in terms of $N$ with variation of $\delta_{1}(t)$. We set unit value for all other parameters.} \label{1}
\end{figure}

\begin{figure}
\centering
\includegraphics[width=0.4\textwidth]{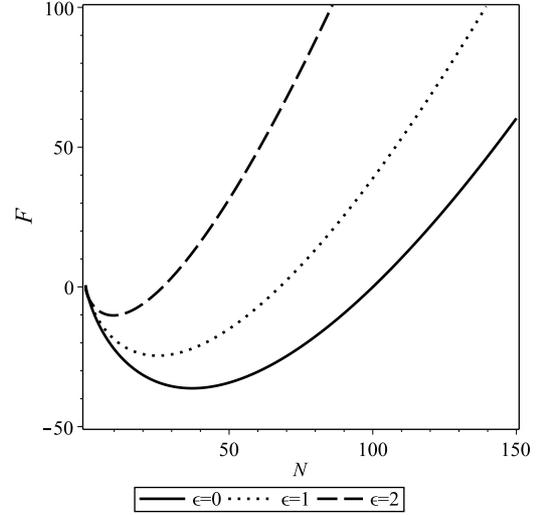}\\
~~~\\
~~~\\
~~~\\
~~~\\
~~~\\\caption{Typical behavior of the Helmholtz free energy in terms of $N$ with variation of $\epsilon$. Black lines are corresponding to modified gravity while red lines are corresponding to standard general relativity. We set unit value for all other parameters.}
\label{2}
\end{figure}

Our goal of this section is to study the effects of $f(R)$ gravity on various thermodynamical
equations of state of the theory.  Once we have explicit expression for
gravitational partition function is known, it is matter of calculation to derive various
thermodynamical entities.
For instance, with the help of relation between  partition function
and  Helmholtz free energy as $F=-T\ln Z_{N}(T,V)$,
we can derive  Helmholtz free energy  as
\begin{equation}
F=-T\ln\biggl(\frac{1}{N!}\left(\frac{2\pi MT}{\lambda^2}\right)^{3N/2}V^N\big(1+\alpha x\big)^{N-1}\biggr).
\end{equation}
This further simplifies to
\begin{eqnarray}
F&=&NT\ln\left(\frac{N}{V}T^{-3/2}\right)-NT -(N-1)T\ln\big(1+\alpha x\big)
\nonumber\\
& -&\frac{3}{2}NT\ln\left(\frac{2\pi M}{\lambda^2}\right).\label{f}
\end{eqnarray}
In the Fig. \ref{1} and Fig. \ref{2} we can see behavior of the Helmholtz free energy in terms of $N$ with variation of $\delta_{1}(t)$ (Fig. \ref{1}) and $\epsilon$ (Fig. \ref{2}). It is illustrated that there is a minimum for Helmholtz free energy. We show that both parameter enhanced Helmholtz free energy. It means that modified gravity correction  increases value of the Helmholtz free energy. For the enough large $\delta_{1}(t)$ the value of the Helmholtz free energy is positive. Hence, if $\delta_{1}(t)$ be increasing function of time, hence Helmholtz free energy is increasing function of time (later with analyze of the entropy we show that $\delta_{1}(t)$ must be increasing function of time). We also find that Helmholtz free energy corresponding to point masses is smaller than Helmholtz free energy corresponding to the extended masses. Moreover, by the red lines of the Fig. \ref{2} (lower lines) we can see that effects of modified gravity (comparison with standard general relativity) is increasing of Helmholtz free energy.
This point deserves some comments.  As we have shown, the Helmholtz free energy is a quantity  depending on $\alpha$, and $\alpha$ depends on $\epsilon$.  As $\epsilon$ is a parameter  introduced to take into account  extended structures, the Helmholtz free energy results different  with respect to  that of point-like  masses. In our specific case, improving $\epsilon$ gives rise to the improvement of the Helmholtz free energy. From a physical point of view,
this means that the Helmholtz free energy  depends  on the size of the system (the size of the galaxy)  and the interaction potential, (here a Newtonian potential corrected with a Yukawa term). Being such a free energy an {\it extensive} thermodynamical quantity, depending on the volume, it is obviously larger than that of a point-like system.

Since entropy $S$ and Helmholtz free energy are related with the following  formula:
$S= -\biggl(\frac{\partial F}{\partial T}\biggr)_{N,V}$. Hence, corresponding to Helmholtz free energy (\ref{f}), the entropy is given by,
\begin{eqnarray}
S&=&N\ln\left(\frac{V}{N}T^{3/2}\right)+(N-1)\ln\big(1+\alpha x\big)\nonumber\\
&-&3N\frac{\alpha x}{1+\alpha x}+\frac{5}{2}N+\frac{3}{2}N\ln\left(\frac{2\pi M}{\lambda^2}\right).\label{s}
\end{eqnarray}
Here we note that  $N$ is large enough to assume  $N-1\approx N$.
This leads to following specific entropy:
\begin{equation}\label{s2}
\frac{S}{N}=\frac{S_{0}}{N}-3\frac{\alpha x}{1+\alpha x}+\ln\left(\frac{V}{N}T^{3/2}\right)-\ln\big(1-\frac{\alpha x}{1+\alpha x}\big),
\end{equation}
where  $S_{0}=\frac{5}{2}N+\frac{3}{2}N\ln\left(\frac{2\pi M}{\lambda^2}\right)$.
These expressions coincide with their   standard form \citep{ahm02}
except the modification in the form of clustering parameter of galaxies,  $\mathcal{B}$, defined as
\begin{eqnarray}\label{b}
\mathcal{B}=\frac{\alpha x}{1+\alpha x}.
\end{eqnarray}
Here, we note that the form of the modified clustering parameter emerges naturally.
The clustering parameter is an important quantity because its value governs the strength of
clustering.

\begin{figure}
\centering
\includegraphics[width=0.4\textwidth]{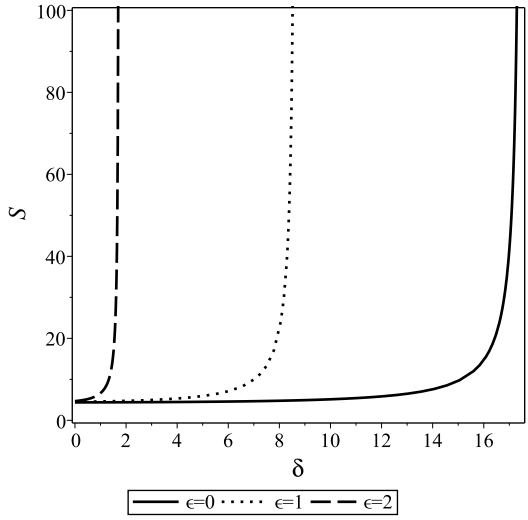}\\
~~~\\
~~~\\
~~~\\
~~~\\
~~~\\\caption{Typical behavior of the entropy in terms of $\delta_{1}(t)$ by variation of $\epsilon$. We set unit value for all other parameters.}
\label{3}
\end{figure}

In case of point masses, i.e.  $\epsilon =0$,  the modified clustering parameter under $f(R)$ gravity takes the following form:
\begin{eqnarray}
\mathcal{B}(\epsilon =0):=\mathcal{B}_0  =\frac{(1+\alpha_2|_{\epsilon=0}) x}{1+x(1+
\alpha_2 |_{\epsilon=0})}.
\end{eqnarray}
In the Fig. \ref{3} we can see behavior of the entropy with modified gravity parameter $\delta_{1}(t)$. It is clear that if $\delta_{1}(t)$ be increasing function of time, then the entropy is also increasing function of time in agreement with the second law of the thermodynamics. We also can see that the entropy of the point masses diverges after the entropy of the extended masses. In order to have comparison with standard general relativity one can take $\xi\rightarrow\infty$ limit and obtain red (three lines in right) lines of the Fig. \ref{3}. We can see similar behavior in general but with different asymptotic points. Also in the case of $\delta_{1}(t)\rightarrow0$ both have similar behavior as expected.

\begin{figure}
\centering
\includegraphics[width=0.4\textwidth]{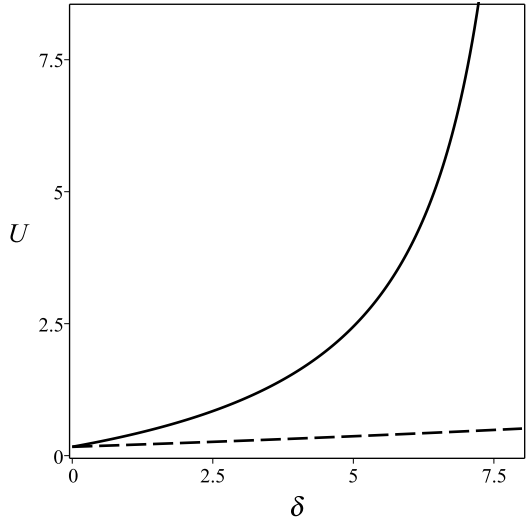}\\
~~~\\
~~~\\
~~~\\
~~~\\
~~~\\\caption{Typical behavior of the internal energy in terms of $\delta_{1}(t)$ for modified gravity (solid line) and standard general relativity (dashed line) We set unit value for all parameters.}
\label{3-1}
\end{figure}

Since the explicit forms of Helmholtz free energy and entropy are known, therefore
 the form of internal energy, $U$,  can be easily derived with the help of definition, $U =  F+TS$, as
\begin{equation}
U=\frac{3}{2}NT\big(1-2\mathcal{B}\big).
\end{equation}
The form of internal energy coincides with the standard form.
However, corrections due to $f(R)$ gravity is embedded in   clustering parameter $\mathcal{B}$. In the Fig. \ref{3-1} we can see general behavior of the internal energy in the modified gravity (solid line) and standard general relativity (dashed line). We can see that the effect of modified gravity is increasing of the internal energy.

The pressure, $P$ and chemical potential, $\mu$, are related with Helmholtz free energy
as
$P= -\left(\frac{\partial F}{\partial V}\right)_{N,T}$ and  $\mu = \biggl(\frac{\partial F}{\partial N}\biggr)_{V,T}$, respectively. Therefore,  the explicit form of pressure  and chemical potential can be given as
\begin{eqnarray}
  P&=&\frac{NT}{V}\big(1-\mathcal{B}\big),\label{p}\\
  {\mu}&=&{T}\ln\left(\frac{N}{V} T^{-3/2}\right)+{T}\ln\big(1-\mathcal{B}\big)\nonumber\\
  &-&\frac{3}{2}{T}\ln\left(\frac{2\pi M}{\lambda^2}\right)-\mathcal{B}{T}.
\end{eqnarray}
These expressions also coincide with their standard forms. The only difference is the
value of clustering parameter which exhibits correction due to the $f(R)$ gravity.
 This chemical potential can be utilized to study the distribution function of the system. We find that both $P$ and $\mu$ behave similar internal energy hence the effect of modified gravity is increasing of the pressure and chemical potential.
\section{General Distribution Function}
\begin{figure}
\centering
\includegraphics[width=0.4\textwidth]{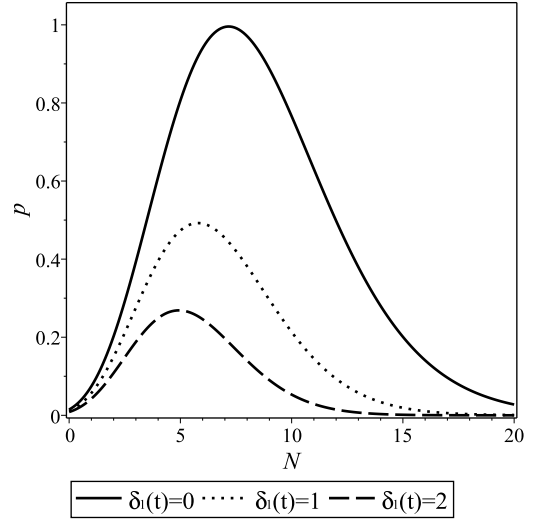}\\
~~~\\
~~~\\
~~~\\
~~~\\
\caption{Typical behavior of the distribution function in terms of $N$ by variation of $\delta_{1}(t)$. We set unit value for all other parameters.}
\label{4}
\end{figure}

Here, we assume that the system of galaxy is in quasi-equilibrium and follows
 grand canonical ensemble.
For grand canonical ensemble, the  partition function is defined by
\begin{eqnarray}
Z_{G}(T,V,z)=\sum_{N=0}^{\infty}z^NZ_{N}(V,T), \label{g}
\end{eqnarray}
where $z$ refers to absolute activity.
The  grand partition function for the system of galaxies  interacting through $f(R)$ gravity is given by
\begin{equation}
\ln Z_{G}=\frac{PV}{T}=  N(1-\mathcal{B}).\label{z}
\end{equation}
For a statistical system, the distribution function (probability of finding $N$ particles in volume $V$)  is given by
\begin{eqnarray}
p_{V}(N)=\frac{\sum_{i}e^{\frac{N\mu}{T}}e^{\frac{-U_i}{T}}}{Z_{G}(T,V,z)}=\frac{e^{\frac{N\mu}{T}}Z_{N}(V,T)}{Z_{G}(T,V,z)}.
\end{eqnarray}
Exploiting relation  (\ref{z}), the distribution function   of gravitational system with extended mass interacting under $f(R)$ gravity is given by
\begin{eqnarray}\label{p000}
&&p_{V}(N,\epsilon)=\frac{\bar{N}^{N}}{N!}\biggl(1+\frac{N}{\bar N}\frac{\mathcal{B}}{(1-\mathcal{B})}\biggr)^{N-1}\nonumber\\
&&\times\biggl(1+\frac{\mathcal{B}}{(1-\mathcal{B})}\biggr)^{-N}e^{[-N\mathcal{B} -\bar N(1-\mathcal{B})]}.
\end{eqnarray}
For point mass case, i.e. $\epsilon=0$, the expression of distribution function  reads,
\begin{eqnarray}
&&p_{V}(N,\epsilon =0)=\frac{\bar{N}^{N}}{N!}\biggl(1+\frac{N}{\bar N}\frac{\mathcal{B}_0}{(1-\mathcal{B}_0)}\biggr)^{N-1}
\nonumber\\
&&\times \biggl(1+\frac{\mathcal{B}_0}{(1-\mathcal{B}_0)}\biggr)^{-N}e^{[-N\mathcal{B}_0 -\bar N(1-\mathcal{B}_0)]}.
\end{eqnarray}
From the above expression, it is evident that, beside the expression of clustering parameter $\mathcal{B}$, the form of distribution function exactly matches  with their standard form given in Ref. \citep{ahm02}.  The form of
standard clustering parameter, $b$, is given by,
\begin{equation}\label{b000}
b=\frac{x}{1+x},
\end{equation}
which is clearly different than the our clustering parameter  (\ref{b}). In the Fig. \ref{4} we draw distribution function (\ref{p000}) in terms of $N$ and see that probability of finding $N$ particles in volume $V$ in the modified gravity $\delta_{1}(t)\neq0$ is smaller than the case of $\delta_{1}(t)=0$. It is the case because total volume in modified gravity is bigger than the case in standard general relativity, and it is clear for example from the equation (\ref{xxxx}).\\
Easily by taking $\xi\rightarrow\infty$ limit one can obtain standard general relativity probability which is higher than the case of modified gravity.

\section{Critical Temperature}
In this section, we study the stability and critical temperature  of the galaxy clustering.
The  specific heat and their variation for perfect gas   to
virialized gas  ($\mathcal{B} = 1$) have illuminating physical insights into clustering.
The expression of specific heat at constant volume is given by
\begin{eqnarray}
C_V =\frac{1}{N}\left(\frac{\partial U}{\partial T}\right)_{N,V}=\frac{3}{2}\left[\frac{1+6\alpha x -4\alpha^2 x^2}{(1+\alpha x)^2}   \right].
\end{eqnarray}
It can be clearly seen from the expression that in the limit  $\alpha \rightarrow 0$ (and hence $\mathcal{B}\rightarrow0$), the specific heat at constant volume reduces to $C_V=3/2$. This means that in this limit system behaves a monotonic perfect gas system.
On the other hand,  for completely virialized system where galaxies are strongly clustered, (i.e., $\alpha x \rightarrow \infty$ or $\mathcal{B} = 1$), the
  specific heat takes following value:
\begin{eqnarray}
C_V =  - \frac{3}{2}.
\end{eqnarray}
The negative value of specific heat signifies that the system is  unstable.
Here, we note that the instability of system of galaxies are not analogous to the system of imperfect gases as gravitating systems contribute extra degrees
of freedom due to semistability. The negative specific heat can be justified as
 suppose many galaxies attach to clusters and therefore  impart energy to clusters which
causes galaxies to rise out of cluster potential wells. Consequently, galaxies  lose kinetic energy and become cooler, therefore
 producing the negative overall value of specific heat.
Moreover, in limit  $\alpha x \rightarrow \frac{1}{2}$, the specific heat of the system reduces to
  \begin{eqnarray}
C_V =   \frac{5}{2},
\end{eqnarray}
which shows the behavior of diatomic gas.
 By
maximizing $C_V$ with temperature, i.e.,
\begin{eqnarray}
\frac{\partial C_V}{\partial T} =0,
\end{eqnarray}
we can get the value of critical temperature.

The critical temperature for point structured galaxies  is obtained as
\begin{eqnarray}
T_{c}(\epsilon=0) =\left[ {3\frac{\bar N}{V}(GM^2)^3} ({1+\alpha_2|_{\epsilon=0}}) \right]^{1/3}.
\end{eqnarray}
For extended structure galaxies, the expression of critical temperature is
 \begin{eqnarray}
T_c =\left[ {3\frac{\bar N}{V} (GM^2)^3}({\alpha_1+\alpha_2})
\right]^{1/3},
\end{eqnarray}
where parameter $\alpha_{1}$ and $\alpha_{2}$ have explicit expressions in (\ref{alp}).

The specific heat for both point and extended mass cases can be expressed in terms of critical temperature as
\begin{eqnarray}
C_V =\frac{3}{2}\left[1-2\frac{1-4\left(\frac{T}{T_c}\right)^3}{\left(1+2\left(\frac{T}{T_c}\right)^3 \right)^2}\right].
\end{eqnarray}
The above form coincides exactly with the the form given in Ref.  \citep{5aa}.
This justifies  the claim even in $f(R)$ case that at critical temperature
the basic homogeneity of the
system has been broken on the average interparticle scale
by the formation of binary systems  bounded gravitationally.
We express the clustering parameter in terms of critical temperature as
\begin{eqnarray}
\mathcal{B}=\frac{T_c^3}{T_c^3+2 {T}^3 }.
\end{eqnarray}
This implies  that, at critical temperature, i.e. $T=T_c$, the   corrected clustering parameter takes following value: $\mathcal{B}_{crit}=1/3$. We can call this as a critical value of clustering parameter
at which the specific heat takes its maximum value.

The  pressure and internal energy can be expressed  in terms of critical temperature as following:
 \begin{eqnarray}
&& P=\frac{2N}{V}\left(\frac{T^4  }{ T_c^3+2  {T}^3 }\right),\\
&&{U}= \frac{3}{2}NT\left(\frac{2T^3-T_c^3  }{ 2T^3+T_c^3  }\right).
 \end{eqnarray}
It is evident that at critical temperature, i.e. $T=T_c$, the   pressure and internal energy take following values:
 $P=\frac{2}{3}\frac{NT}{V}$ and ${U}= \frac{1}{2}NT$, respectively.

\section{The galaxy-galaxy correlation function in F(R) gravity}
In this section, we would like to examine the behavior of two particle correlation function
under the effect of $F(R)$ gravity. In fact, the power law behavior of
two particle correlation function $\Xi(x)$ in clustering of galaxies, i.e.  $\Xi(x)=r^{-1.8}$, has been established by observation \citep{pee80} as well as by N-body simulation \citep{sut90}.

The internal energy $U$ for a  system of spherical volume $V$ consisting of $N$-particles has
following form \citep{sas00}:
\begin{equation}
U=\frac{3}{2}NT-\frac{N\rho}{2}\int_{V}\Phi(r)\Xi(r)4\pi r^2dr,
\end{equation}
where interaction potential between two galaxies  $\Phi$ in $f(R)$ gravity is given in Eq. (\ref{12}).
Now, in order to see the modification in the power law of correlation function,
 the clustering parameter is defined as
\begin{equation}
\mathcal{B}=\frac{GM^2\rho}{6T}\int\frac{r\Xi(r)}{\sqrt{\epsilon^2 +r^2}}\big(\frac{1}{f_1r}+\frac{\delta_1(t)e^{-r\sqrt{-\xi}}}{6GM\xi}\big)dV.
\end{equation}
The volume differentiation of this yields following expression:
\begin{eqnarray}
&&\frac{\partial \mathcal{B}}{\partial V}=\frac{Gm^2\rho}{6T}\frac{d\rho}{dV}   \int\frac{r\Xi(r)}{\sqrt{\epsilon^2 +r^2}} \big(\frac{1}{f_1r}+\frac{\delta_1(t)e^{-r\sqrt{-\xi}}}{6GM\xi}\big)  dV   \nonumber \\
&&+ \frac{Gm^2\rho}{6T}\frac{\partial}{\partial V}\int\frac{r\Xi(r)}{\sqrt{\epsilon^2 +r^2}}\big(\frac{1}{f_1r}
+\frac{\delta_1(t)e^{-r\sqrt{-\xi}}}{6GM\xi}\big)dV.\label{ss}
\end{eqnarray}
Inserting following  relation
\begin{equation}
\frac{\partial \mathcal{B}}{\partial V}=\frac{\partial \rho}{\partial V}\frac{\partial \mathcal{B}}{\partial \rho}
\end{equation}
in  Eq. (\ref{ss}), we get
\begin{equation}
\frac{\partial \mathcal{B}}{\partial \rho}=\frac{\mathcal{B}(1-\mathcal{B})}{\rho}.
\end{equation}
Since $\rho=\frac{N}{V}$, therefore
\begin{equation}
\frac{\partial \rho}{\partial V}=-\frac{\rho}{V}.
\end{equation}
Upon further simplifications of above relations, we get the following functional form of correlation function:
\begin{equation}
\xi(r)=\frac{9\mathcal{B}^2T}{2\pi Gm^2\rho}\frac{1}{r^2}\bigl(1+\frac{\epsilon^2}{2r^2}\bigr)\bigl( \frac{1}{f_1r}+\frac{\delta_1(t)e^{-r\sqrt{-\xi}}}{6GM\xi} \bigr)^{-1},\label{funtt}
\end{equation}
where we have neglected the higher powers of $\epsilon/r$.
From relation (\ref{funtt}), the effects of $f(R)$ modifications to the correlation function can be easily seen. First of all, we notice how $f_1$ would only change the amount of correlation (the larger $f_1$, the larger $\xi$), but not the radial behavior; instead, $\delta_1$ has a twofold effect, both on the degree of correlation (the larger $\delta_1$, the smaller $\xi$) and on the radial dependence through the exponential behaviour. Finally, as explained in the previous sections, it is not easy to quantify and convert present constraints on $f(R)$ in terms of changes in the correlation function, due to the variety of $f(R)$ models which are considered, and to the chosen theoretical parameters.

\section{Observational data}

{\renewcommand{\tabcolsep}{1.5mm}
{\renewcommand{\arraystretch}{2.}
\begin{table*}
\begin{minipage}{\textwidth}
\caption{Results from fitting Eq.~(\ref{p000}) with data from the cluster catalog from \citep{WenHan2012}. Data and fits are divided in redshift bins (first column) and cell radius bin (as indicated by $R$ intervals). $N_{cl}$ is not a fitting parameter, but the number of clusters in each redshift and radius bin. All the fitting parameters are described in the text.}\label{tab:results}
\centering
\resizebox*{\textwidth}{!}{
\begin{tabular}{c|cccccc|cccccc|cccccc}
\multicolumn{19}{c}{$\chi^{2}_{ls}$} \\
\hline
 & \multicolumn{6}{c|}{$0.8<R_1<0.9$ Mpc} & \multicolumn{6}{c|}{$0.9<R_1<1.0$ Mpc} & \multicolumn{6}{c}{$1.0<R_1<1.1$ Mpc}  \\
$z$          & $N_{cl}$ & $\bar{N}$                & $b$           & $\tilde{L}$ & $\tilde{\delta}_1$ & $f_1$
             & $N_{cl}$ & $\bar{N}$ & $b$ & $\tilde{L}$ & $\tilde{\delta}_1$ & $f_1$
             & $N_{cl}$ & $\bar{N}$ & $b$ & $\tilde{L}$ & $\tilde{\delta}_1$ & $f_1$ \\
\hline
$[0.05,0.1]$ &  $291$ & $9.33^{+0.12}_{-0.13}$ & $<0.16$ & $<9.16$ & $<3.75$ & $\mathit{290}$ &
$390$ & $11.16^{+0.19}_{-0.20}$ & $<0.22$ & $<4.43$ & $<6.38$ & $\mathit{2955}$ &
$298$ & $14.04^{+0.09}_{-0.09}$ & $<0.10$ & $<1.11$ & $\mathit{162}$ & $\mathit{4.99}$ \\
$[0.1,0.15]$ & $1052$ & $10.01^{+0.15}_{-0.15}$ & $0.069^{+0.056}_{-0.056}$ & $<1.20$ & $<125$ & $\mathit{191}$ &
$1526$ & $11.96^{+0.10}_{-0.10}$ & $<0.16$ & $<2.14$ & $<28.1$ & $\mathit{0.33}$ &
$949$ & $15.59^{+0.05}_{-0.05}$ & $<0.19$ & $<2.96$ & $<2.44$ & $\mathit{60.9}$ \\
$[0.15,0.2]$ & $2343$ & $10.54^{+0.16}_{-0.17}$ & $<0.23$ & $<4.01$ & $<11.1$ & $\mathit{18}$ &
$2919$ & $12.93^{+0.11}_{-0.10}$ & $<0.21$ & $<5.71$ & $<3.12$ & $\mathit{97.3}$ &
$1826$ & $16.75^{+0.03}_{-0.03}$ & $0.05^{+0.20}_{-0.04}$ & $0.47^{+0.44}_{-0.27}$ & $<18.6$ & $\mathit{53.6}$ \\
$[0.2,0.25]$ & $3069$ & $10.58^{+0.14}_{-0.14}$ & $<0.26$ & $<4.91$ & $<6.43$ & $\mathit{4219}$ &
$3688$ & $12.85^{+0.05}_{-0.05}$ & $<0.094$ & $<1.24$ & $<67.8$ & $\mathit{165}$ &
$2398$ & $16.92^{+0.07}_{-0.07}$ & $0.076^{+0.069}_{-0.057}$ & $<0.93$ & $\mathit{25.7}$ & $\mathit{335}$ \\
$[0.25,0.3]$ & $3318$ & $10.13^{+0.14}_{-0.14}$ & $<0.16$ & $<1.68$ & $<70.6$ & $\mathit{2716}$ &
$3982$ & $12.56^{+0.15}_{-0.15}$ & $<0.25$ & $<4.21$ & $<4.93$ & $\mathit{27.2}$ &
$2443$ & $16.17^{+0.10}_{-0.10}$ & $<0.16$ & $<4.33$ & $<5.71$ & $\mathit{88.9}$ \\
$[0.3,0.35]$ & $4024$ & $10.28^{+0.14}_{-0.14}$ & $<0.096$ & $<1.36$ & $<148$ & $\mathit{905}$ &
$5055$ & $12.48^{+0.15}_{-0.15}$ & $<0.13$ & $<1.39$ & $<56.5$ & $\mathit{182}$ &
$2958$ & $16.35^{+0.13}_{-0.13}$ & $<0.20$ & $<2.82$ & $<8.17$ & $\mathit{39.9}$ \\
$[0.35,0.4]$ & $4379$ & $10.36^{+0.14}_{-0.14}$ & $<0.22$ & $<6.63$ & $<4.95$ & $\mathit{0.26}$ &
$5402$ & $12.70^{+0.16}_{-0.15}$ & $<0.098$ & $<1.01$ & $\mathit{15.0}$ & $\mathit{90.0}$ &
$3208$ & $16.37^{+0.13}_{-0.13}$ & $<0.19$ & $<2.10$ & $<11.8$ & $\mathit{63.5}$ \\
$[0.4,0.45]$ & $4925$ & $10.09^{+0.18}_{-0.18}$ & $<0.20$ & $<4.27$ & $<16.1$ & $\mathit{13.5}$ &
$5819$ & $12.35^{+0.19}_{-0.18}$ & $<0.13$ & $<1.21$ & $<96.3$ & $\mathit{17771}$ &
$3289$ & $16.10^{+0.19}_{-0.19}$ & $<0.13$ & $<2.59$ & $<26.9$ & $\mathit{218}$ \\
$[0.45,0.5]$ & $3982$ & $9.45^{+0.11}_{-0.12}$ & $<0.15$ & $<0.59$ & $<216$ & $\mathit{554}$ &
$4813$ & $11.38^{+0.17}_{-0.17}$ & $<0.23$ & $<3.83$ & $<12.5$ & $\mathit{634}$ &
$2660$ & $14.65^{+0.17}_{-0.17}$ & $<0.08$ & $<0.81$ & $<171$ & $\mathit{131}$ \\
$[0.5,0.55]$ & $2838$ & $8.92^{+0.22}_{-0.21}$ & $<0.18$ & $<2.91$ & $<34.3$ & $\mathit{35.9}$ &
$3631$ & $10.15^{+0.14}_{-0.14}$ & $<0.16$ & $<1.36$ & $\mathit{47.8}$ & $\mathit{5450}$ &
$2009$ & $12.98^{+0.15}_{-0.16}$ & $<0.19$ & $<2.65$ & $<19.5$ & $\mathit{1594}$ \\
$[0.55,0.6]$ & $1971$ & $8.69^{+0.24}_{-0.25}$ & $<0.11$ & $<1.48$ & $<201$ & $\mathit{3.39}$ &
$2614$ & $9.61^{+0.34}_{-0.36}$ & $<0.23$ & $<3.99$ & $<14.1$ & $\mathit{0.19}$ &
$1538$ & $12.28^{+0.15}_{-0.15}$ & $<0.20$ & $<3.60$ & $<8.23$ & $\mathit{119}$ \\
$[0.6,0.65]$ & $1046$ & $8.49^{+0.20}_{-0.21}$ & $<0.29$ & $<4.47$ & $<8.30$ & $\mathit{215}$ &
$1586$ & $9.26^{+0.32}_{-0.34}$ & $<0.18$ & $<1.99$ & $<63.3$ & $\mathit{908}$ &
$977$ & $11.18^{+0.18}_{-0.18}$ & $<0.15$ & $<3.67$ & $<20.7$ & $\mathit{18.9}$ \\
\hline
\multicolumn{19}{c}{$\chi^{2}_{jk}$} \\
\hline
 & \multicolumn{6}{c|}{$0.8<R_1<0.9$ Mpc} & \multicolumn{6}{c|}{$0.9<R_1<1.0$ Mpc} & \multicolumn{6}{c}{$1.0<R_1<1.1$ Mpc}  \\
$z$          & $N_{cl}$ & $\bar{N}$                & $b$           & $\tilde{L}$ & $\tilde{\delta}_1$ & $f_1$
             & $N_{cl}$ & $\bar{N}$ & $b$ & $\tilde{L}$ & $\tilde{\delta}_1$ & $f_1$
             & $N_{cl}$ & $\bar{N}$ & $b$ & $\tilde{L}$ & $\tilde{\delta}_1$ & $f_1$ \\
\hline
$[0.05,0.1]$ &  $291$ & $9.39^{+0.08}_{-0.08}$ & $<0.20$ & $<1.95$ & $<36.6$ & $\mathit{9.6\cdot10^5}$ &
$390$ & $11.26^{+0.07}_{-0.07}$ & $0.04^{+0.10}_{-0.03}$ & $0.48^{+0.49}_{-0.30}$ & $<204$ & $\mathit{2791}$ &
$298$ & $14.26^{+0.08}_{-0.08}$ & $<0.13$ & $0.63^{+1.13}_{-0.42}$ & $<8.57$ & $\mathit{7.0\cdot10^5}$ \\
$[0.1,0.15]$ & $1052$ & $10.15^{+0.04}_{-0.05}$ & $0.13^{+0.15}_{-0.08}$ & $<0.78$ & $<57.8$ & $\mathit{117}$ &
$1526$ & $12.01^{+0.04}_{-0.04}$ & $<0.15$ & $<0.98$ & $<35.6$ & $\mathit{1156}$ &
$949$ & $15.55^{+0.04}_{-0.04}$ & $<0.10$ & $<0.95$ & $<17.9$ & $\mathit{7.4\cdot10^4}$ \\
$[0.15,0.2]$ & $2343$ & $10.69^{+0.03}_{-0.03}$ & $<0.24$ & $<1.39$ & $<25.1$ & $\mathit{2106}$ &
$2919$ & $12.99^{+0.03}_{-0.03}$ & $<0.11$ & $<1.02$ & $<62.1$ & $\mathit{31.7}$ &
$1826$ & $16.81^{+0.04}_{-0.03}$ & $<0.06$ & $<0.75$ & $<21.3$ & $\mathit{342}$ \\
$[0.2,0.25]$ & $3069$ & $10.57^{+0.07}_{-0.07}$ & $<0.10$ & $<1.04$ & $<103$ & $\mathit{156}$ &
$3688$ & $12.97^{+0.03}_{-0.03}$ & $<0.16$ & $<0.99$ & $<29.6$ & $\mathit{705}$ &
$2398$ & $16.94^{+0.03}_{-0.03}$ & $<0.11$ & $<0.77$ & $<27.9$ & $\mathit{2673}$ \\
$[0.25,0.3]$ & $3318$ & $10.26^{+0.03}_{-0.03}$ & $0.14^{+0.21}_{-0.09}$ & $<0.84$ & $<61.2$ & $\mathit{956}$ &
$3982$ & $12.59^{+0.03}_{-0.03}$ & $<0.16$ & $<0.92$ & $<55.1$ & $\mathit{8.71}$ &
$2443$ & $16.19^{+0.03}_{-0.03}$ & $<0.08$ & $<0.99$ & $<50.4$ & $\mathit{152}$ \\
$[0.3,0.35]$ & $4024$ & $10.44^{+0.03}_{-0.03}$ & $<0.25$ & $<1.81$ & $<24.0$ & $\mathit{116}$ &
$5055$ & $12.55^{+0.03}_{-0.03}$ & $<0.30$ & $<1.82$ & $<8.21$ & $\mathit{816}$ &
$2958$ & $16.38^{+0.03}_{-0.03}$ & $<0.20$ & $<0.68$ & $<29.8$ & $\mathit{146}$ \\
$[0.35,0.4]$ & $4379$ & $10.52^{+0.03}_{-0.03}$ & $0.12^{+0.14}_{-0.07}$ & $<0.72$ & $<74.3$ & $\mathit{63.6}$ &
$5402$ & $12.75^{+0.02}_{-0.02}$ & $<0.05$ & $2.73^{+4.62}_{1.43}$ & $<19.3$ & $\mathit{41.1}$ &
$3208$ & $16.45^{+0.03}_{-0.03}$ & $<0.19$ & $<0.86$ & $<23.0$ & $\mathit{11.6}$ \\
$[0.4,0.45]$ & $4925$ & $10.31^{+0.038}_{-0.03}$ & $<0.29$ & $<3.08$ & $\mathit{23.4}$ & $\mathit{54.8}$ &
$5819$ & $12.45^{+0.02}_{-0.03}$ & $<0.11$ & $<0.94$ & $<99.7$ & $\mathit{1067}$ &
$3289$ & $16.08^{+0.03}_{-0.03}$ & $0.12^{+0.16}_{-0.07}$ & $<0.87$ & $\mathit{3.30}$ & $\mathit{14.6}$ \\
$[0.45,0.5]$ & $3982$ & $9.78^{+0.03}_{-0.03}$ & $<0.35$ & $<3.69$ & $\mathit{14.2}$ & $\mathit{176}$ &
$4813$ & $11.50^{+0.03}_{-0.03}$ & $<0.19$ & $<1.02$ & $<50.8$ & $\mathit{817}$ &
$2660$ & $14.69^{+0.03}_{-0.03}$ & $0.15^{+0.20}_{-0.10}$ & $<0.93$ & $<25.6$ & $\mathit{131}$ \\
$[0.5,0.55]$ & $2838$ & $9.25^{+0.03}_{-0.03}$ & $<0.26$ & $4.66^{+5.22}_{-2.43}$ & $<2.27$ & $\mathit{0.86}$ &
$3631$ & $10.33^{+0.03}_{-0.03}$ & $0.32^{+0.21}_{-0.15}$ & $<0.99$ & $<15.9$ & $\mathit{52.7}$ &
$2009$ & $13.0^{+0.03}_{-0.03}$ & $<0.17$ & $<0.82$ & $<48.7$ & $\mathit{1201}$ \\
$[0.55,0.6]$ & $1971$ & $9.06^{+0.04}_{-0.04}$ & $<0.25$ & $<3.45$ & $<12.2$ & $\mathit{2.08}$ &
$2614$ & $10.37^{+0.05}_{-0.05}$ & $<0.28$ & $<1.13$ & $<36.5$ & $\mathit{2443}$ &
$1538$ & $12.38^{+0.05}_{-0.05}$ & $<0.11$ & $<1.07$ & $<53.9$ & $\mathit{158}$ \\
$[0.6,0.65]$ & $1046$ & $8.85^{+0.05}_{-0.05}$ & $<0.24$ & $<1.54$ & $\mathit{1.84}$ & $\mathit{1.4\cdot10^4}$ &
$1586$ & $9.50^{+0.05}_{-0.04}$ & $<0.19$ & $<1.20$ & $<73.6$ & $\mathit{41.3}$ &
$977$ & $11.32^{+0.05}_{-0.05}$ & $<0.19$ & $<1.12$ & $<37.2$ & $\mathit{61.5}$ \\
\hline
\end{tabular}}
\end{minipage}
\end{table*}}}

In order to test the feasibility of an $f(R)$-gravity approach to the clustering of galaxies using some of the tools developed in the previous sections, we need to choose a galaxy and/or cluster catalog to be compared with theory. We have chosen to work with the cluster catalog described in \citep{WenHan2012}, containing $132,684$ clusters of galaxies in the redshift range $0.05 \leq z < 0.8$. The catalogue is based on the identification of clusters and groups of galaxies from the Sloan Digital Sky Survey III (SDSS-III). Even if there is a more updated version of the same catalogue \citep{WenHan2015}, we have decided to work with this older version because it provides, for each cluster, all the quantities we need to test our theory by using Eq.~(\ref{p000}). In particular, among others, the catalogue provides:
\begin{itemize}
 \item the radius of the cluster, $r_{200}$ in $Mpc$, which is, as usual, the radius within which the mean density of a cluster is $200$ times the critical density of the universe at the same redshift;
 \item $N_{200}$, the number of member galaxy candidates for each given cluster, within $r_{200}$.
\end{itemize}
The most recent version of the same catalogue \citep{WenHan2015} provides slightly more accurate estimations for clusters' masses, but at $r_{500}<r_{200}$; as we are not interested in masses, but only in galaxy counts-in-cells, and we want to explore the largest volumes possible, we will stick to the older numbers from \citep{WenHan2012}, even if a comparison between the clustering at $r_{500}$ and at $r_{200}$ might be interesting. But we will leave it for future works.

How do we perform the counts-in-cell procedure? First, we identify the cells with the volumes corresponding to each $r_{200}$; thus, for us, each cluster in the catalogue is a properly defined cell. Then, for each cluster/cell, we derive the observed distribution function $p_{V}(N)$ by counting how many galaxies are in each cell, i.e., by using the given $N_{200}$ data. Considering that we are given both the redshift $z$ and the radius $r_{200}$ for each cluster, we divide the full sample in smaller groups and try to detect, if any, a possible dependence of the theoretical parameters with scale and time. We have divided the total sample in redshift bins with width $\Delta z = 0.05$, and radius bins with width $\Delta r_{200} = 0.1$ $Mpc$. From the obtained sub-groups, we have selected only those containing a sufficiently large number of clusters, ending with the groups corresponding to cell/cluster radii $0.8 < r_{200} < 0.9$, $0.9 < r_{200} < 1.0$ and $1.0 < r_{200} < 1.1$ $Mpc$, and to redshift $0.05 \leq z \leq 0.65$. We need to remember that the catalogue we are using is complete up to a redshift $z \sim 0.42$, for clusters with an estimated mass $M_{200}> 1 \cdot 10^{14}$ $M_{\odot}$; at higher redshift, a bias toward smaller clusters (lower masses) is possible \citep{WenHan2012}.

It is important to highlight here some main differences between the way we perform our counts-in-cells analysis and previous works in literature. In most of the latter ones \citep{4a,Sivakoff}, once a catalogue is chosen, the covered sky is divided in cells by fixing their angular dimensions, and the galaxy count is performed in each cell for each chosen angle. These counts are then compared with the theoretical expectation for the distribution function $p_{V}(N)$. Instead, we follow more closely the approach of \citep{Yang11,Yang12}, where the cells are determined by their physical length, in our case the radius $r_{200}$. This latter approach has also another benefit which becomes more important and evident when working with a sample which extends on a large redshift interval, like in our case: the same physical lenght is used as reference cell radius at every redshift, while the former approach, with the cells being determined by constant angular dimensions, would imply that larger physical cells/volumes are explored at higher redshift with respect to lower redshift ones.

Another maybe even more important difference, however, is in the way the cells are defined. All the former references start from a galaxy sample, define a cell volume, and cover the full sky by this mapping; then,  galaxies are counted in each cell but without taking into consideration the \textit{clustering status} of the galaxies, i.e., if they form structures (groups and clusters) or if they are field galaxies. What is generally argued, is that the cell radius, at least numerically, \textit{could} correspond to a cluster and/or group of galaxies scale, but there is no certainty that such structures are present and no correspondence is looked for concerning this aspect. In \citep{WenHan2012}, the galaxy catalog from SDSS-III is scanned exactly for this purpose, namely, to search for groups and clusters of galaxies, and is eventually returned as a cluster catalog. Thus, we will explore the compatibility of the theoretical distribution function with consistently-identified clustered structures. This same analysis has been started for the first time in \citep{Yang12}, but on a much smaller redshift range than that we consider in this work. Also, as we are working on scales which are much smaller than those where the quasi-equilibrium clustering should be more effective, there will be some consequences and caveats which we will have to take in mind when interpreting our results, as explained in \citep{Yang11,Yang12}.

In Table~\ref{tab:results} we report results from fitting Eq.~(\ref{p000}) with the chosen cluster catalog data. The parameters involved in Eq.~(\ref{p000}) are rewritten as:
\begin{itemize}
  \item[-] $R_1=r_{200}$ is the volume radius of the cell (cluster) we have considered; as explained above, we bin the data with respect to $R_1$;
  \item[-] $N=N_{200}$ is the number of galaxies in each cell/cluster;
  \item[-] the dimensionless softening parameter:
        \begin{equation}
        \tilde{\epsilon} = \frac{\epsilon}{R_1}\; ;
        \end{equation}
  \item[-] the interaction range of the $f(R)$ model, both dimensional (in $Mpc$):
        \begin{equation}
        L = \frac{1}{\sqrt{-\xi}}= \sqrt{-\frac{6f_{2}}{f_{1}}} \; ;
        \end{equation}
        and dimensionless:
        \begin{equation}
        \tilde{L} = \frac{L}{R_1} \; .
        \end{equation}
        Thus, it is clear we leave $\tilde{L}$ and $f_{1}$ as free parameters in our fitting procedure, while $f_{2}$ can be determined from them;
  \item[-] the clustering parameter $b$, which enters Eq.~(\ref{p000}) through:
        \begin{equation}
        \mathcal{B} = \frac{b\, \alpha}{1+b(\alpha-1)} \; ;
        \end{equation}
  \item[-] the dimensionless time-function parameter which quantify the deviation of the $f(R)$-model from General Relativity (GR):
        \begin{equation}
        \tilde{\delta_{1}} = \frac{\delta_{1}\, R_{1}^{2}}{G\,M}\, .
        \end{equation}
\end{itemize}
Note that in GR we have only two parameters for Eq.~(\ref{p000}), i.e. $\bar{N}$ and $b$ which might be even derived directly from observations, if the counts-in-cells were performed on scales large enough for the quasi-equilibrium condition to hold. Thus, in principle, in GR, we would have no free parameters. However, it is also clear that if one wants to perform a fit of the same data with a given expression for the distribution function, a maximum of two free parameters are present. Then, it is obvious that for our $f(R)$ model, which has five free parameters (including the $f(R)$ parameters $\tilde{L},\tilde{\delta_{1}},f_{1}$), and given the way they enter in Eq.~(\ref{p000}), we might have a lot of degeneracies in our case and very likely we won't be able to constrain in a statistically satisfying way most of the parameters. In Table~\ref{tab:results} we also include $N_{cl}$, i.e. the number of clusters in each redshift bin, but this won't be a fitting parameter.

\begin{figure*}
\includegraphics[width=\textwidth]{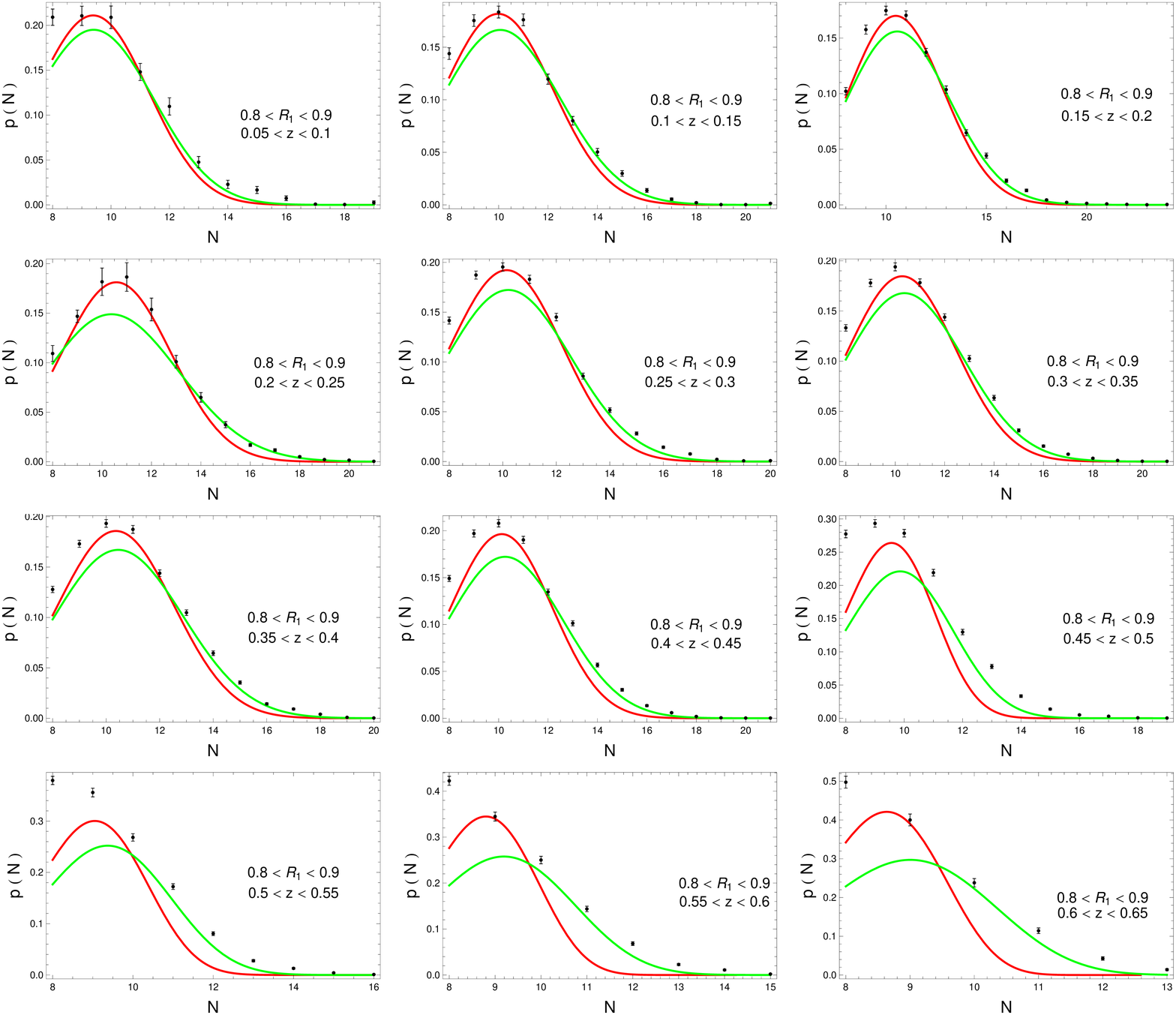}
\caption{Comparison between data and theoretical expectation Eq.~(\ref{p000}) for cell size bin $0.8 < R_{1} < 0.9$ $Mpc$. Black points: data; black bars: jackknife-like observational errors. Solid red line: best fit from minimization of $\chi_{ls}^{2}$; solid green line: best fit from minimization of $\chi_{jk}^{2}$.}\label{5}
\end{figure*}

\begin{figure*}
\includegraphics[width=\textwidth]{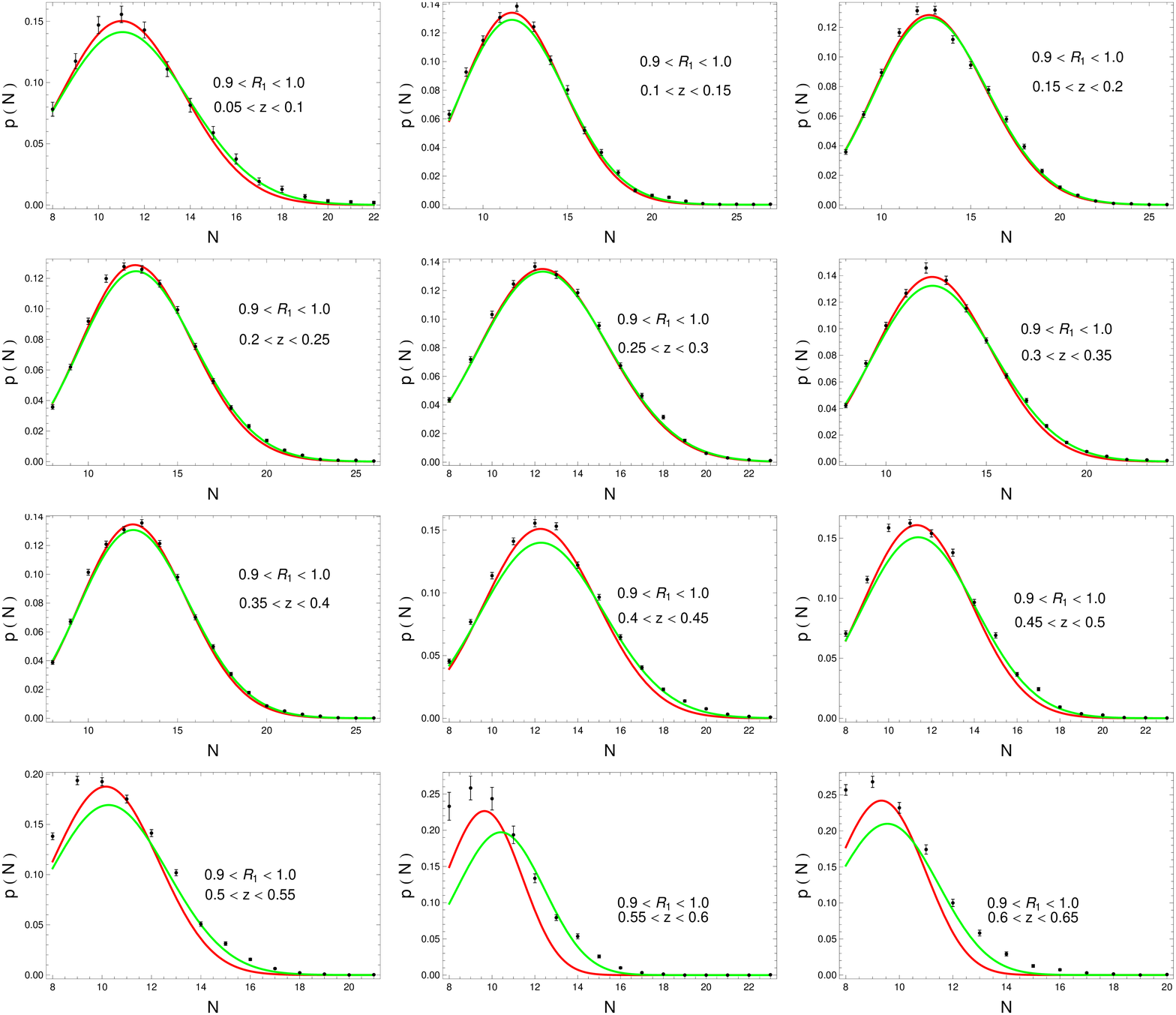}
\caption{Comparison between data and theoretical expectation Eq.~(\ref{p000}) for cell size bin $0.9 < R_{1} < 1.0$ $Mpc$. Black points: data; black bars: jackknife-like observational errors. Solid red line: best fit from minimization of $\chi_{ls}^{2}$; solid green line: best fit from minimization of $\chi_{jk}^{2}$.}\label{6}
\end{figure*}

\begin{figure*}
\includegraphics[width=\textwidth]{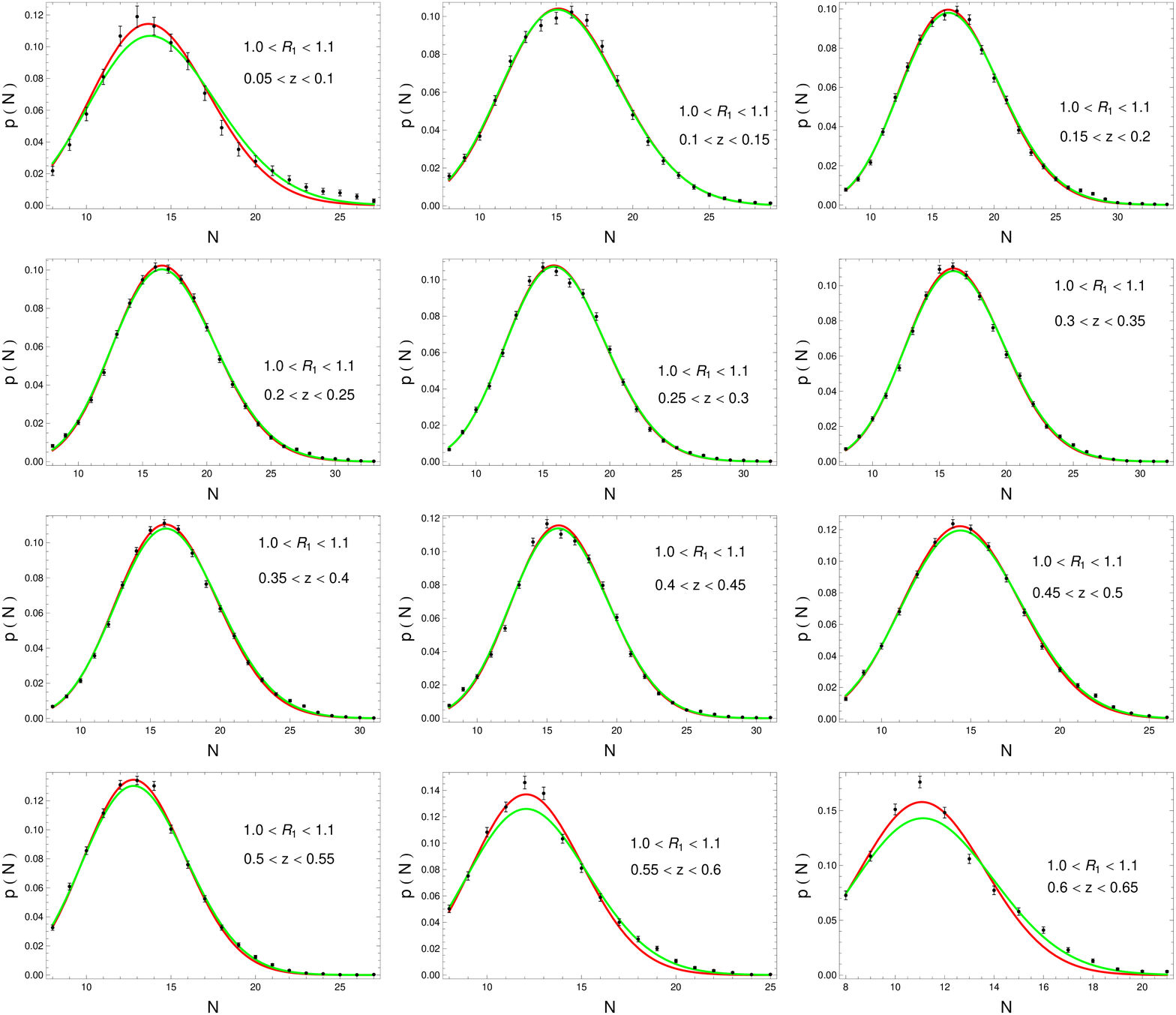}
\caption{Comparison between data and theoretical expectation Eq.~(\ref{p000}) for cell size bin $1.0 < R_{1} < 1.1$ $Mpc$. Black points: data; black bars: jackknife-like observational errors. Solid red line: best fit from minimization of $\chi_{ls}^{2}$; solid green line: best fit from minimization of $\chi_{jk}^{2}$.}\label{7}
\end{figure*}

In order to fit the data, we have used two different approached. First, we have performed a proper least-square minimization, i.e. we minimize the quantity
\begin{equation}
\chi^{2}_{ls} = \sum_{i}\left(p^{theo}_{V}(N_{i}) - p^{obs}_{V}(N_{i})\right)^2,
\end{equation}
where $p^{obs}_{V}(N_{i})$ is the counts-in-cell distribution from data, and $p^{theo}_{V}(N_{i})$ is the counts-in-cell expected from $f(R)$, as given in Eq.~(\ref{p000}). The index $i$ derives from the fact that in each bin, we have a minimal and maximum $N_{200}$; we select $i$ finite values in the range $[N_{min},N_{max}]$ at which we evaluate the distribution function, both observational and theoretical. Second, we also perform a $\chi^2$-like analysis, minimizing the quantity:
\begin{equation}
\chi^{2}_{jk} = \sum_{i} \frac{\left(p^{theo}_{V}(N_{i}) - p^{obs}_{V}(N_{i})\right)^2}{\sigma^{2}_{i}},
\end{equation}
where $\sigma_i$ are the errors on $p^{obs}_{V}(N_{i})$. We have derived these errors from the data with a jackknife-like procedure. For each redshift and radius bin:
\begin{enumerate}
  \item we cut a fraction $\mathcal{F}$ from the total bin population;
  \item we derive the counts-in-cell distribution function for the cut sample, i.e. we have a new distribution for the $N_i$ defined as above.
\end{enumerate}
The previous steps are performed for different fraction $\mathcal{F}$, ranging from $10\%$ to $90\%$ of the total sample; and for each fraction, we repeat the same steps up to $\sim 50$ times. In such a way, we end up with a very-closely-gaussian distribution of $\sim 50$ $N_i$ (for each $i$) values; for each of them, we extract the standard deviation and assume this last quantity as the error $\sigma_i$. The data points and the errors so obtained are shown as black dots and bars in Figs.~(\ref{5})~-~(\ref{6})~-~(\ref{7}). The best fitting $p_{V}(N)$ distributions obtained from the minimization of $\chi^2_{ls}$ are shown in red; those derived from $\chi^2_{jk}$ are in green.

The minimization of the defined $\chi^2$ is performed by using a Monte Carlo Markov Chain (MCMC) approach, running chains with $10^6$ points and applying some priors on the parameters in the most general way possible, namely: $\bar{N}\geq0$; $0\leq b \leq 1$; $0\leq \tilde{\epsilon} \leq 1$; $\tilde{L}>0$; $f_{1}>0$, given that the GR limit is $f_{1}\rightarrow 1$ and we need the first term in Eq. (5) to have attractive behaviour; and $\tilde{\delta}_{1}>0$. This last condition ensures that the $f(R)$ model we are considering works as a dark energy-type fluid, i.e. gives repulsive contribution to the gravitational potential. We could have considered also the case $\delta_1<0$, implying an attractive contribution to the gravitational potential and, thus, having the $f(R)$ theory mimicking a dark matter effect. However, such effect is quite controversial and questionable (see, for example, \citep{Sotiriou} and references therein), and would require a deeper theoretical study which is of course out of the purpose of this work. Thus, we stick to the only choice of a positive $\delta_1$.

Moving the discussion to the constraints we are able to put on the theoretical parameters, as it can be easily noted, in Table~\ref{tab:results} we do not report results for the softening parameter $\tilde{\epsilon}$, which is definitely unconstrained, showing a basically uniform distribution all over the feasible range $[0,1]$. On the other hand, as expected, $\bar{N}$ is very well constrained in all cases, and there is no statistically significant difference between the values obtained with $\chi^2_{ls}$ and $\chi^2_{jk}$. The clustering parameter $b$ is mostly consistent with zero, and only an upper limit can be set; this result indicates that the clusters in the catalog are not virialized \citep{Yang12}. Moreover, a similar trend has been noted also in literature, where smaller values of the cells correspond to smaller values of $b$ \citep{Sivakoff,4a,Yang12}.

The interaction length $\tilde{L}$ (which is a parameter characteristic of the gravitational theory considered), as well as $b$, is basically consistent with zero which is, incidentally, its GR limit. To be noted that $\tilde{L}$ depends on $f_{1}$ and $f_{2}$, which both enter the action of the $f(R)$ model and, thus, should be fixed by the theory to well determined values valid for all the gravitational structures once and for all. Instead, we have left them free to vary from one cluster to another in order to take into account possible local interactions leading the clusters far from the quasi-equilibrium condition. If we join the probability distributions from each and every cluster, we end up with a final joint estimation for the full sample, of $\tilde{L} = 0.47 \pm 0.09$ $Mpc$ for $\chi_{ls}^{2}$ and of $\tilde{L} = 0.41 \pm 0.08$ $Mpc$ for $\chi_{jk}^{2}$. The two are very consistent with each other, but we cannot conclude that we do detect a deviation from GR, because we miss to check two other parameters, $\tilde{\delta}_{1}$ and $f_{1}$.

And, in fact, the $f(R)$-amplitude parameter $\tilde{\delta}_{1}$ itself is mostly consistent with zero, and only an upper limit can be set in most of the cases. In terms of the dimensional parameter $\delta_1$, assuming an average galaxy mass $10^{13}$ $M_{\odot}$ (in \citep{Ahmad06} it has been shown that a mass range has secondary effects which are mostly negligible, so that to assume identical masses for galaxies, as it is usually done, is a good approximation), we easily find $\delta < \mathcal{O}(10^2)$. While it might seem a large signal, one must not forget it is coupled to an exponential term which drops fast its contribution. Moreover, while a time dependence is explicitly theoretically allowed for it, the data return no evidence for a possible evolution. Finally, the parameter $f_{1}$ is totally unconstrained. Note that for $f_{1}$ and for some cases of $\tilde{\delta}_{1}$ we have not given proper confidence intervals as best fit estimations, but we have only indicated some values in italic font, which correspond to the values assumed by the parameters exactly in the minimum of the two $\chi^2$ previously defined. When this happens, it is because no statistics can be derived, due to a very noisy and irregular distribution. Note that for $f_{1}$ the values corresponding to the minimum in the $\chi^2$ are somehow not physical (because of the degeneracies): very high values of $f_{1}$ would make the standard Newtonian part in the gravitational potential much smaller than what should be expected with respect to the new correction terms from the $f(R)$ gravity modifications.

Anyway, all this is not strange at all, due to very large degeneracies among all the parameters involved in the analysis. First of all, the quantity $\alpha_1$ in Eq.~(19) is strongly correlated with the value of $f_{1}$, see Eq.~(23); if $\tilde{\epsilon}$ is basically unconstrained, so it is $\alpha_1$ and, as result, $f_{1}$ too. The impact of $\tilde{\epsilon}$ on $\alpha_2$, from Eq.~(19), is smaller, but it is easy to see that $\alpha_2 \propto \tilde{\delta}_{1} \tilde{L}$, so that we have another possible degeneracy between these two parameters. Anyway, as said, both $\tilde{\delta}_{1}$ and $\tilde{L}$ (when single bin results are considered) are consistent with zero which is the limit for GR to apply, together with $f_{1} \rightarrow 1$; but we are not in condition to check if it satisfied or not, due to the bad statistics related to this parameter.

As more general considerations, we can note how high redshift clusters are fitted worse, in general, than smaller redshift ones; this could be due to the uncomplete sampling from the catalog. The same reason might very likely be at the base of the trend in $\bar{N}$, which rises slightly with redshift, becomes practically constant, and then starts to decrease for $z>0.45$. As we have explained above, the sample we are using is complete up to $z \sim 0.42$, after which we can have a bias toward smaller, namely, less massive, clusters (i.e. with smaller $\bar{N}$).

Finally, we also note that the larger is the cell volume (larger is $R_{1}$), the better is the fit. The bins $0.8 < R_{1} < 0.9$ $Mpc$ give unsatisfactory fits with both $\chi^2$ definitions, at every redshift interval. In general, smaller redshifts have a better agreement, with $\chi_{ls}^2$ fitting better small $N$ values and around the peak in the distribution function; while $\chi_{jk}^{2}$ gives better fit at high $N$. Cells slightly larger, with $0.9 < R_{1} < 1.0$ $Mpc$, show a much better agreement between theory and data up to $z \sim 0.4$, with any difference between $\chi_{ls}^2$ and $\chi_{jk}^2$ barely distinguishable. At higher redshifts $0.4 < z < 0.55$, we fit data different, depending on the $\chi^2$, as described in the previous point; while data at very high redshifts $z>0.55$ are not fit by the theoretical distribution. The largest cells, with $1.0 < R_{1} < 1.1$ $Mpc$, instead, show remarkable agreement between theory and data, independently of the $\chi^2$ definition used, up to $z\sim 0.55$, with slightly larger inconsistencies at higher redshifts, even if not so prominent as in previous cases.

\section{Discussion and Conclusions}
In this paper, we have studied the clustering of galaxies under the $f(R)$ gravity model.
By assuming that the clustering of galaxies  in an expanding universe evolves in a
quasi-equilibrium manner through a sequence of equilibrium states, we have evaluated the
gravitational partition function. With the help of resulting partition function, we have
estimated the various equations of states of the such system where effects of
$f(R)$ modifications are evident. In particular, we have derived explicit expressions for
Helmholtz free energy, entropy, internal energy,  pressure and chemical potential.
The effects of $f(R)$ modifications on Helmholtz free energy and entropy are
also analysed with the help of plot.  We have found that modified gravity
correction  increases  the value of the Helmholtz free energy. From the plot, we have seen that if modified gravity parameter $\delta_1(t)$ be increasing
function of time, then the entropy is also increasing
function of time, which is in agreement with the second law of
the thermodynamics.
Interestingly, we have found that  the expressions of equations of states match with their
standard expression except the form of clustering parameter which emerges naturally. The clustering parameter exhibits the effects of $f(R)$ modification.
Since the system follows a quasi-equilibrium state,
this enforces us to assume the system as a grand canonical ensemble.
With this assumption, we  have calculated the exact distribution function  for the system
which depends upon $f(R)$ corrected clustering parameter.  We have discussed the critical temperature and
conditions of stability also for the system. The resulting expression of specific heat obtained in  case of $f(R)$ gravity  justified the claim that at critical temperature the basic homogeneity of the system has been broken on the average inter-particle scale by the formation of binary systems bounded gravitationally. We have obtained a relation between the clustering parameter and critical temperature, which estimates the  critical value of clustering parameter at which the specific heat takes its maximum value. The effects of $f(R)$ modifications on the two-particle correlation function is also studied.

The feasibility of the gravitational quasi-equilibrium described in a thermodynamical way from the $f(R)$-model has been also tested with observations. In particular we have made use of the distribution function of gravitational systems with extended mass interacting under $f(R)$ gravity given by Eq.~(\ref{p000}), and describing how many galaxies are clustered/clustering in a spatial volume cell/region. We have used the cluster catalog described in \citep{WenHan2012}, which provides the number of galaxies belonging to each of the $\sim 10^5$ clusters identified from the SDSS-III survey, together with their redshift and size. We have divided the full sample in spatial (by cluster's size) and time (by redshift) bins, under the condition they contain an enough large number of objects in order to have a good statistical description of the underlying distribution function, to verify the possibility of a space-time dependence of the same function. All these information make us able to try a fit of our theoretical results with real data. In order to perform the counts-in-cell analysis, we have divided the total sample in redshift bins with width $\Delta z = 0.05$, and radius bins with width $\Delta r_{200} = 0.1$ $Mpc$. From the obtained sub-groups, we have selected only those containing a sufficiently large number of clusters, ending with the groups corresponding to cell/cluster radii $0.8 < r_{200} < 0.9$, $0.9 < r_{200} < 1.0$ and $1.0 < r_{200} < 1.1$ $Mpc$, and to redshift $0.05 \leq z \leq 0.65$. By doing so, we must remember that the considered catalogue  is complete up to a redshift $z \sim 0.42$, for clusters with an estimated mass $M_{200}> 1 \cdot 10^{14}$ $M_{\odot}$; at higher redshift, a bias toward smaller clusters (lower masses) is possible \citep{WenHan2012}.

On one hand, as shown in Figs.~(5)~-~(6)~-~(7), the fit is quite remarkably good, except for a regime (high redshift tail and lower cluster size), for which a departure from the quasi-equilibrium description is visible. Very likely, this depends on the not-completeness of the used catalog in such regime; but also it is known from literature that the quasi-equilibrium is better verified (and more easily detectable) at scales much larger than the cluster's ones we are studying here. Anyway, as it is possible to see from Fig.~(7), even for such strongly clustered systems, but with larger dimensions, we have a very good agreement between theory and observations.

On the other hand, we have too many free parameters entering in our analysis, and a lot of degeneracy among most of them, so that many have very poor constraints. This seems to be a problem of the $f(R)$ theory, more than the approach itself. In fact, the two parameters which are in common with GR, i.e. $\bar{N}$ and $b$ (the minimal number of parameters required to describe the clustering of galaxies in the thermodynamical quasi-equilibrium ), are very well constrained and consistent with other results in literature. In particular, the very small values for $b$, which indicates a farer from equilibrium state of the cluster with the sizes we have considered. All the other parameters are more weakly constrained: the softening parameter $\tilde{\epsilon}$ (which is not a gravity theory parameter) is basically unconstrained; the dimensionless interaction length of the $f(R)$ theory, $\tilde{L}$, is the best constrained parameter, but it is consistent with the GR value $(\tilde{L} \rightarrow 0)$, and only an upper limit can be given for most of the clusters from the sample; the deviation from GR, $\tilde{\delta}_{1}$, is consistent with the GR limit, too, namely, $(\tilde{\delta}_{1} \rightarrow 0)$, but the related statistics is somehow invalidated by the last parameter, $f_{1}$, which is always unconstrained and with very irregular confidence contours. In some way, $f_{1}$ is the elected capstone of all the degeneracies among all the parameters of our model.

In fact, we also have found that the fit is much better for the larger volume cells. The bins $0.8 < R_{1} < 0.9$ $Mpc$ give unsatisfactory fits with both $\chi^2$ definitions, at every redshift interval. In general, smaller redshifts have a better agreement, with $\chi_{ls}^2$ fitting better small $N$ values and around the peak in the distribution function; while $\chi_{jk}^{2}$ gives better fit at high $N$. Cells slightly larger, with $0.9 < R_{1} < 1.0$ $Mpc$, show a much better agreement between theory and data up to $z \sim 0.4$, with any difference between $\chi_{ls}^2$ and $\chi_{jk}^2$ barely distinguishable. At higher redshifts $0.4 < z < 0.55$,  depending on the $\chi^2$, data fits  different; while data at very high redshifts $z>0.55$ are not fit by the theoretical distribution. Interestingly, the largest cells, with $1.0 < R_{1} < 1.1$ $Mpc$, instead, show remarkable agreement between theory and data, independently of the $\chi^2$ definition used, up to $z\sim 0.55$, with slightly larger inconsistencies at higher redshifts.

Thus, we conclude that the distribution function from the quasi-equilibrium approach might be used to constrain theories only if the number of additional parameters is small, and the correlation/degeneracy among them is strongly reduced. Then, we plan to use such approach with other than GR gravity models with such requirements.

\section*{Acknowledgements}

V.S. is financed by the Polish National Science Center Grant DEC-2012/06/A/ST2/00395. Contribution from S.C. and V.S. is also based upon work from COST action CA15117 (CANTATA), supported by COST (European Cooperation in Science and Technology).

\bsp
\label{lastpage}
\end{document}